# Coding Scheme for Optimizing Random I/O Performance

Eran Sharon, Idan Alrod
SanDisk Corporation

**Abstract:** Flash memories intended for SSD and mobile applications need to provide high random I/O performance. This requires using efficient schemes for reading small chunks of data (e.g. 0.5KB – 4KB) from random addresses. Furthermore, in order to be cost efficient, it is desirable to use high density Multi-Level Cell (MLC) memories, such as the ones based on 3 or 4 bit per cell technologies. Unfortunately, these two requirements are contradicting, as reading an MLC memory, whose data is coded conventionally, requires multiple sensing operations, resulting in slow reading and degraded random I/O performance. This paper describes a novel coding scheme that optimizes random read throughput, by allowing reading small data chunks from an MLC memory using a single sensing operation.

## A. Background

Flash memories intended for SSD and mobile applications need to provide high random I/O performance. This requires using efficient schemes for programming and reading small chunks of data (e.g. 0.5KB – 4KB) from random addresses (e.g. aligned to 0.5KB). Furthermore, in order to be cost efficient, it is desirable to use high density Multi-Level Cell (MLC) memories, such as the ones based on 3 or 4 bit per cell technologies. Unfortunately, these two requirements are contradicting, as high density memories tend to have slow programming and reading and this poses a problem for high random I/O performance.

Conventionally, each physical page, corresponding to a WordLine (WL), in an MLC Flash stores $log_2(M)$ logical pages of data, where $M$ is the number of levels to which each cell is programmed. E.g. a WL of a 4-bit-per-cell Flash stores 4 logical pages, by programming each cell to 16 levels.

In order to allow fast reading of an MLC Flash, one needs to minimize the number of sense operations that are needed in order to retrieve the required data, wherein each sensing operation compares all cells of the WL to a single voltage threshold and determines for each cell whether its voltage level is smaller than the threshold. In that sense, an interleaved coding scheme, in which the data is encoded into a codeword which is spanned over all the logical pages of the WordLine (WL), are inefficient. The reason is that they require $M-1$ sense operations for retrieving any data chunk (even 0.5KB), as the data is encoded on all $M$ states. More efficient coding schemes for optimizing random read performance are non-interleaved (or semi-interleaved) coding schemes. In these coding schemes the basic data chunk (e.g. 0.5KB or 1KB) is encoded over a single logical page, allowing one to retrieve it using only $(M-1)/log_2(M)$ sense operations on the average (i.e. faster by a factor of $log_2(M)$ compared to an interleaved coding scheme).

**Can we do better?** This paper describes a coding scheme that optimizes random read throughput, by allowing reading of small data chunks using only a single sensing operation. I.e. faster by a factor of $M-1$ compared to an interleaved coding scheme and faster by a factor of $(M-1)/log_2(M)$ compared to non-interleaved and semi-interleaved coding schemes. E.g. assume an 4-bit-per-cell Flash example as shown in Figure 1, an interleaved coding scheme would require 15 sense operations for reading each data chunk, a non-interleaved coding scheme would require 3.75 sense operations (on average) for reading each data chunk and the proposed coding scheme would require 1 sensing operation for reading each data chunk.

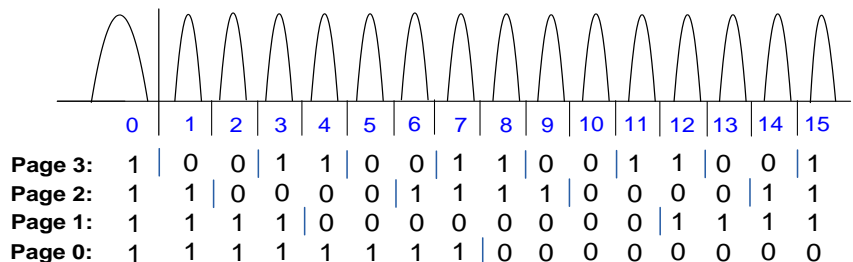

Figure 1: 4-bit-per-cell example

## B. Coding scheme for optimizing random I/O performance

The basic idea of Random I/O (RIO) coding is to divide the data into $M-1$ equal (or possibly non-equal) portions and to encode the $j$'th portion of the data ($j = 1,2,...,M-1$) in the indices of the cells that will be "on" when we perform a single sense operation between the $j-1$ and $j$ programming levels. This way, each $1/(M-1)$ portion of the data can be read using a single sensing operation. More specifically, RIO coding for an MLC Flash with $M$ programming levels encodes the first $1/(M-1)$ portion of the data in the indices of the cells that will be "on" when we perform a sense operation between levels 0 and 1, it encodes the second $1/(M-1)$ portion of the data in the indices of the cells that will be "on" when we perform a sense operation between levels 1 and 2, … and so on, till the $(M-1)$-th portion of the data is encoded in the indices of the cells that will be "on" when we perform a sense operation between levels $M-1$ and $M$.

**How do we know that such a valid encoding exists?**
Consider a Write Once Memory (WOM), a.k.a a One Time Programmable (OTP) memory, in which each cell stores a single bit and can be programmed only once from an erased state ("1") to a programmed state ("0"). Once programmed it cannot change its state anymore. Such a memory can model a punch card, where information is programmed to the card by punching holes in it. It can also model optical OTP media, and Flash OTP media (e.g. EROM). From the work of Rivest and Shamir [1], it is known that there exists a coding scheme that allows to program $k$ bits into $n$ cells of a WOM $t$ times, if $n \geq k*t/\log_2(t)$. We will refer to such a coding scheme as a WOM coding scheme. As proved by Rivest and Shamir, an overhead of $n/k = t/\log_2(t)$ is sufficient in order to ensure the possibility to write $k$ bits into $n$ cells $t$ times. Actually, this overhead is conjectured to be a tight estimate of the minimal required overhead only for large values of $t$. For small values of $t$ it is pessimistic. E.g., for $t = 2$, there exist a coding scheme that requires an overhead of only ~1.2938, which is smaller than $t/\log_2(t) = 2$.

The existence of a valid RIO coding scheme can be proved via a reduction from the problem of multiple programming of a WOM. Consider a coding scheme for writing $k$ bits into $n$ cells of a punch card $t$ times. Let us denote by $I_j$ and $C_j$ the information written in the $j$'th time and the sequence of "holes" punched into the punch card cells in the $j$'th time, respectively. The programming and reading procedure of the punch card is illustrated in Figure 2, for $t = 3$.

Assuming the existence of such WOM coding, we can use the same scheme for implementing the proposed RIO coding for an MLC Flash with $M = t+1$ programming levels. Assume we want to program $K$ bits into the MLC WL. We divide the $K$ bits into $t$ subsets each of size $k = K/t$. We then use the WOM coding scheme to encode each subset of $k$ bits, only that now the "holes" punched into the punch card in the $j$'th programming step ($j = 1,2,...,t=M-1$) are replaced by the label "$M-j$", representing an MLC Flash cell which is programmed to the ($M-j$)-'th level. After we encoded the $t$ subsets of $k$ bits into a sequence of programming levels, cells which were not assigned any programming level so far are assigned with the level "0" (i.e. the Erase level). At this point we have a valid assignment of programming levels to all the cells and we can program the WL accordingly. Now, if we want to read the $j$'th subset of $k$ bits we can perform a single sense operation between programming levels $j-1$ and $j$. The result of this reading is a binary vector, from which we can recover the $k$ bits of information by using the WOM decoding scheme. The programming and reading procedures according to the RIO coding scheme are illustrated in Figure 3, for an MLC Flash with $M = 4$.

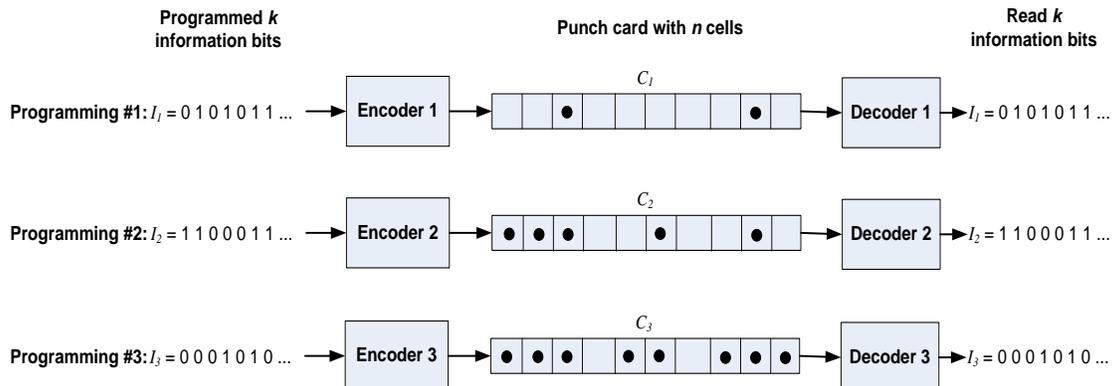

Figure 2: a coding scheme for writing $k$ bits into $n$ cells of a punch card $t=3$ times

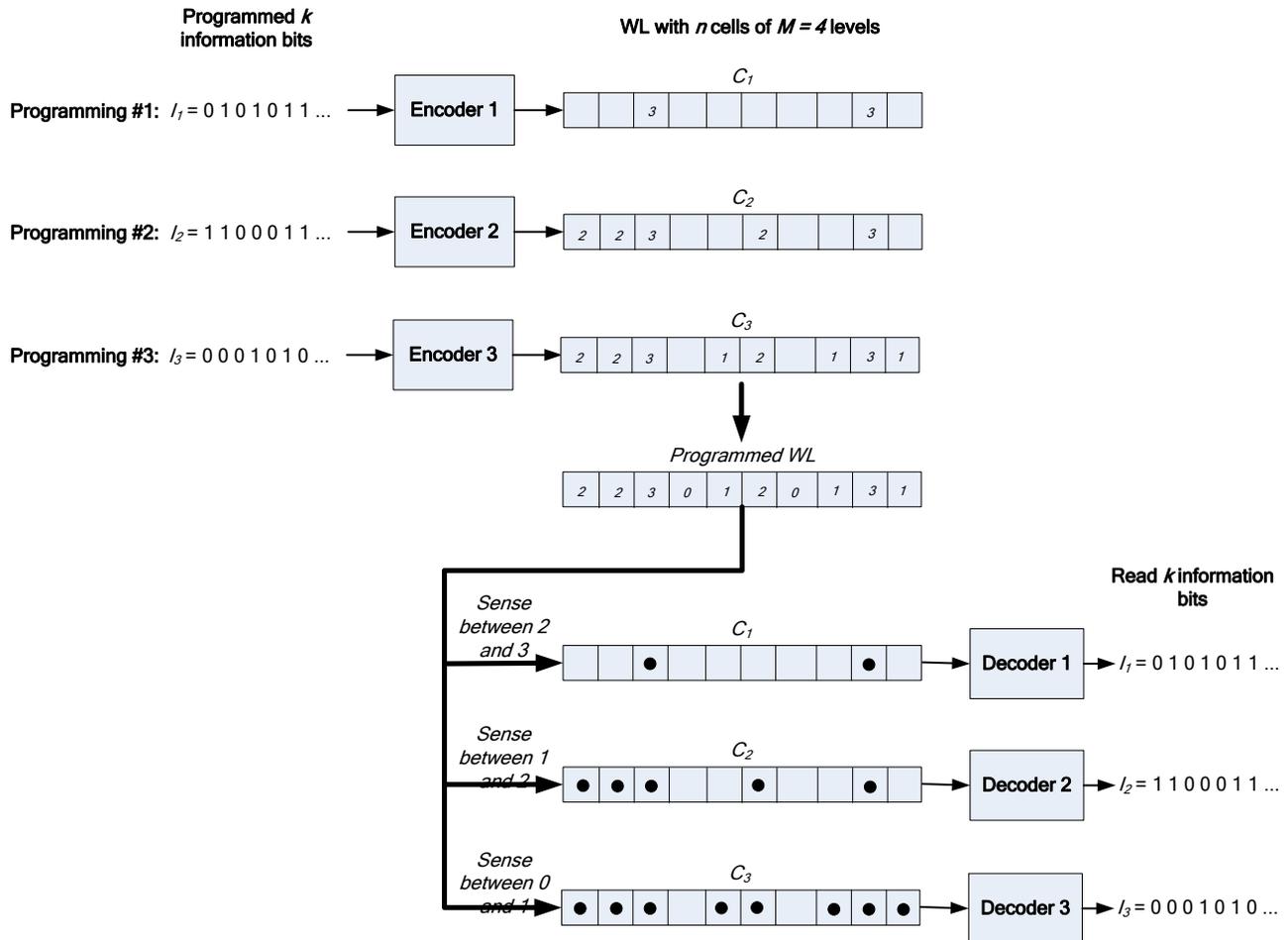

Figure 3: RIO coding scheme for programming $3*k$ bits into $n$ cells of a Flash with $M=t+1=4$

RIO code construction can be based on the same constructions used for WOM ([1,2,3,4]). These schemes are usually based on coset coding, where each information sequence is represented by multiple codewords, such that it is possible to choose a codeword in the current programming step whose support set contains the support set of the codeword programmed in the previous programming step.

A "toy" example of a RIO code is shown below. It is based on a code for programming $k = 2$ bits into $n = 3$ cells of a WOM $t = 2$ times. In the RIO coding context, it is used for allowing programming of $K = k*t = 4$ bits into $n = 3$ MLC cells with $M = t+1 = 3$ programming levels, such that each set of $k = 2$ bits can be recovered using a single sensing operation. The first $k = 2$ bits can be recovered by sensing between levels 2 and 1. The second $k=2$ bits can be recovered by sensing between levels 1 and 0.

The encoding of the "toy" RIO code is defined as follows:

| Second set of information bits | First set of information bits | | | |
|---|---|---|---|---|
| | 0 0 | 0 1 | 1 0 | 1 1 |
| 0 0 | 2 0 0 | 0 2 1 | 0 1 2 | 0 1 1 |
| 0 1 | 2 0 1 | 0 2 0 | 1 0 2 | 1 0 1 |
| 1 0 | 2 1 0 | 1 2 0 | 0 0 2 | 1 1 0 |
| 1 1 | 2 1 1 | 1 2 1 | 1 1 2 | 0 0 0 |

The decoding of the RIO code is defined as follows:

| Read codeword | 011/100 | 010/101 | 001/110 | 000/111 |
|---|---|---|---|---|
| Decoded information | 0 0 | 0 1 | 1 0 | 1 1 |

The "toy" RIO code described above is not efficient as it is very short. For example, when used as a RIO code, it allows programming only $K = k*t = 4$ bits into $n = 3$ MLC cells. I.e. we can store only $4/3 = 1.333$ bits per cell. However, the theoretic capacity of an MLC cell with $M=t+1 = 3$ levels is $log_2(M) = 1.585$ bits per cell. Hence, we are not utilizing the Flash efficiently.
As shown in the next section, an optimal RIO code would allow programming 1.5458 bits per cell.

## C. RIO coding efficiency - asymptotic analysis

In order to understand the achievable efficiency when using an optimal RIO code, we can perform the following asymptotic analysis. We will use the terminology of the problem of multiple programming of a WOM. Let us denote by $p_j$ the fraction of cells that are programmed from "1" to "0" during the $j$'th programming stage (out of the cells that are erased at that point). For sake of simplicity, let's assume that in stage $j$ we use codewords having a fixed number of "0" (i.e. fixed codeword weight), which is $w_j$. This assumption is not limiting asymptotically (as the code length $n$ tends to infinity) because the fraction of length $n$ vectors which have weight of exactly $w_j$ out of the set of length $n$ vectors which have weight $\leq w_j$ tends to 1. This means that by limiting the (asymptotic) analysis to a fixed codeword weight we incur negligible rate penalty (rate penalty tends to 0 as $n$ tends to infinity). During the first programming we would like to use the lowest weight vectors. Clearly, the minimal fraction of cells that has to be programmed during the first stage is given by:

$$2^k = \binom{n}{n \cdot p_1} \cong 2^{nH_b(p_1)} \Rightarrow p_1 = H_b^{-1}(k/n), \quad \text{where } H_b(p) = p \cdot \log_2\left(\frac{1}{p}\right) + (1-p) \cdot \log_2\left(\frac{1}{1-p}\right)$$

Similarly, in the second programming stage, the minimal fraction of erased cells that need to be programmed satisfies:

$$2^k \leq \binom{n \cdot (1-p_1)}{n \cdot (1-p_1) \cdot p_2} \cong 2^{n \cdot (1-p_1) \cdot H_b(p_2)} \Rightarrow p_2 \geq H_b^{-1}\left(\frac{k/n}{(1-p_1)}\right)$$

Similarity, in the $j$'th programming stage, the minimal fraction of erased cells that need to be programmed satisfies:

$$2^k \leq \binom{n \cdot (1-p_1) \cdot (1-p_2) \cdot \ldots \cdot (1-p_{j-1})}{n \cdot (1-p_1) \cdot (1-p_2) \cdot \ldots \cdot (1-p_{j-1}) \cdot p_j} \cong 2^{n \cdot (1-p_1) \cdot (1-p_2) \cdot \ldots \cdot (1-p_{j-1}) \cdot H_b(p_j)}$$

$$\Rightarrow p_j \geq H_b^{-1}\left(\frac{k/n}{(1-p_1) \cdot (1-p_2) \cdot \ldots \cdot (1-p_{j-1})}\right)$$

It is not hard to see that the sequence $p_1,p_2,\ldots$ satisfies $0<p_1\leq p_2\leq\ldots\leq p_j\leq ½$ (as the function $H_b(p)$ is monotonically increasing with $p$ and obtains its maxima at $p = ½$).
Hence, the number of times $t$ that a WOM with $n$ cells can be programmed with $k$ bits is upper bounded by the maximal $j$ for which $(1-p_1) \cdot (1-p_2) \cdot \ldots \cdot (1-p_j) \cdot n \geq k$. It can be shown that this upper bound is achievable [1,5].
Consider for example the case of a RIO coding scheme for $t = 2$. We would like to know what is the minimal overhead $n/k$ which is required for it. From the analysis above we know that we need to satisfy $k \leq n \cdot H_b(p_1)$ and $k \leq (1-p_1) \cdot n$. Hence the minimal required overhead is given by:

$$n/k \geq \min_{p_1}\left\{\frac{1}{H_b(p_1)}, \frac{1}{1-p_1}\right\} = 1.2938 \quad \text{(obtained at } p_1 = 0.2271\text{)}$$

Assuming we use an optimal code and achieve this minimal overhead, how efficient is the RIO coding scheme that we obtain for an MLC with $M=t+1 = 3$ levels?
The optimal scheme allows programming $K = k*t = k*2$ bits into $n$ MLC cells. Hence we can program $2*k/n = 2/1.2938 = 1.5458$ bits per cell. This is much better than what we got with the "toy" code that allowed storing only 1.333 bits per cell. However, it still doesn't achieve the capacity of an MLC cell with $M = 3$ levels, which is

$log2(M)$=1.585 bits per cell. The reason is that the RIO coding scheme introduces inherent "shaping" over the distribution of probabilities to program each programming level, making it non-uniform. The "loss" in storage capability is exactly equal to the redundancy that would be required in order to induce the "shaping" pattern that is introduced by the coding scheme:

According to the analysis above, the optimal values for the fractions of cells that are programmed in each programming stage of the $t = 2$ scheme are $p_1 = 0.2271$ and $p_2 = 0.5$. These values induce the following probability distribution over the $M = 3$ programming levels of the MLC Flash:

$$P = [(1-p_1)\cdot(1-p_2) \quad (1-p_1)\cdot p_2 \quad p_1] = [0.3865 \quad 0.3865 \quad 0.2271]$$

The entropy induced by this probability distribution is

$$H = \sum_{i=1}^{3} P(i) \cdot \log_2\left(\frac{1}{P(i)}\right) = 1.5458$$

as opposed to uniform distribution entropy

$$H = \sum_{i=1}^{3} 1/3 \cdot \log_2\left(\frac{1}{1/3}\right) = \log_2(3) = 1.585$$

I.e. the lost storage capability is exactly equal to the redundancy that is required in order to induce P.

## D. Summary

RIO coding is proposed for allowing reading small chunks of data using a single sensing operation, thus boosting random read performance.

Further research directions may include construction of efficient and low complexity RIO codes and incorporation of RIO coding and Error Correction Coding.